\documentclass[]{emulateapj}
\usepackage[backref, breaklinks, colorlinks, citecolor=blue,linkcolor=magenta]{hyperref}
\usepackage[dvipsnames]{xcolor}

\slugcomment{Submitted to The Astrophysical Journal}

\shorttitle{X-rays in R25: Part II}
\shortauthors{Sanchez et. al.}

\usepackage{natbib}
\begin{document} 

\title{Extra! Extra! X-ray Emission!: \\ X-rays in {\sc Romulus25}, Part II - Physical Drivers of X-ray Emission in the CGM}

\vspace{-2cm}

\author{N. Nicole Sanchez\altaffilmark{1,2}}
\author{Andrew Benson\altaffilmark{1}}
\author{Gwen Rudie\altaffilmark{1}}
\author{Ben Oppenheimer\altaffilmark{3}}
\author{Hannah Bish\altaffilmark{4}}
\author{John ZuHone\altaffilmark{5}}
\author{Michael Tremmel\altaffilmark{6}}


\affil{$^1$The Observatories of the Carnegie Institution for Science, 813 Santa Barbara Street, Pasadena, CA 91101, USA, nsanchez@carnegiescience.edu}
\affil{$^2$Cahill Center for Astronomy and Astrophysics, California Institute of Technology, Pasadena, CA 91125, USA}
\affil{$^3$Departamento de Física Teórica, Módulo 15, Facultad de Ciencias, Universidad Autónoma de Madrid, 28049 Madrid, Spain}
\affil{$^4$European Southern Observatory, Karl-Schwarzschild-Strasse 2, D-85748 Garching bei München, Germany}
\affil{$^5$Center for Astrophysics | Harvard \& Smithsonian, 60 Garden St., Cambridge, MA 02138, USA}
\affil{$^6$School of Physics, University College Cork, Cork, Ireland}
\affil{$^7$Astronomy Department, University of Washington, Box 351580, Seattle, WA, 98195-1580}

\begin{abstract}\label{abs:abstractlabel}

Characterizing the hot (T $\sim$ 10$^{6}$ K) circumgalactic medium (CGM) around Milky Way-mass (MW-mass) galaxies is critical to a full understanding of galaxy evolution. In this second paper of the \textit{X-rays in {\sc Romulus25}} series, we unpack the specific physical processes giving rise to the presence of hot X-ray emitting gas in the halos of galaxies from the {\sc Romulus25} (R25) simulation, focusing on L$_{*}$ galaxies. We determine that a strong correlation appears between the mass of hot CGM gas, M$_{gas, T > 10^{5.5} K}$, and the X-ray emissivity, $\epsilon_{\chi}$, measured in the first paper of this series, (Sanchez et al. 2026a, \textit{submitted}). There is also a secondary correlation between M$_{gas, T > 10^{5.5} K}$ and $\epsilon_{\chi}$ with the average metallicity of the hot CGM gas, $\bar{Z}_{gas, T > 10^{5.5} K}$. We use the well-studied PIONEER simulations from \cite{Sanchez2019}, a set of zoom-in MW-mass galaxies run with the genetic modification technique of \cite{Roth2016} and the same physics as R25, to interpret these correlations. We determine that the presence of hot X-ray emitting gas is determined by two key processes: 1) the quantity of hot gas mass in the CGM, M$_{gas, T > 10^{5.5} K}$, is set by the combined total energy injection of \textit{both} stellar and AGN feedback, and 2) together the average temperature and metallicity of the hot CGM changes the soft X-ray spectrum of the gas resulting in brighter metal-rich CGM and dimmer metal-poor CGM for the same quantity of M$_{gas, T > 10^{5.5} K}$. These results indicate, for the first time, that metal-rich gas halos may bias X-ray observations of MW-mass galaxies.

\end{abstract}


\section{Introduction} 
\label{sec-intro}

Over the last two decades, substantial evidence has come to light showing that a galaxy and its central supermassive black hole (SMBH) co-evolve \citep[][for a review]{Kormendy2013}. 
Studies across the electromagnetic spectrum detect energetic, galaxy-scale outflows being driven from the SMBH. SMBH-driven winds have been shown to inject energy and momentum (feedback) into their surrounding medium via direct radiation \citep{Ciotti2010}, relativistic jets of plasma \citep{Fabian2012}, synchrotron emission in radio \citep{Padovani2017, Panessa2019}, and galaxy scale multi-phase gas winds  \citep[observed in UV and X-rays][and references therein]{Crenshaw2003, Crenshaw2012, Mestici2024}.  These observational constraints have been crucial to developing our understanding of the impact of feedback energy on gas dynamics in galaxy halos. 

Despite significant observational evidence for the SMBH-driven feedback impacting galaxies across mass scales \citep[for recent reviews see][]{Kormendy2013,Alexander2025}, the process through which SMBH feedback energy couples to gas remains uncertain. Many simulation groups are actively working to explain this process and clarify our understanding of the gas phase-AGN feedback connection, with a recent focus on the gas in the circumgalactic medium \citep{Suresh2017,Nelson2018,Oppenheimer2020,Truong2020,Truong2021,davies_galaxy_2022,Truong2023,Grayson2025,Piacitelli2025,Piacitelli2026}.

In recent years, the CGM has been considered a critical laboratory for understanding feedback processes particularly through the analysis of hot gas. Comparisons between simulations and UV spectroscopic observations have illuminated the ways that the gas phase and metallicity of galaxy CGM across stellar mass ranges are being impacted by feedback from SMBHs and stars. Hydrodynamic simulations have predicted that SMBHs heat gas in the CGM, evacuate gas out of the CGM through galaxy-scale winds \citep{Oppenheimer2020,davies_galaxy_2022}, and enrich the CGM by ejecting metals (formed in the disk) out into the halo \citep{Nelson2018,Sanchez2019,Sanchez2024}.

More recently, the launch of the Spectrum Roentgen Gamma (SRG) X-ray space observatory and the subsequent releases of eROSITA all-sky survey data, have provided new observations testing current models, such as new measurements from the hotter phases of gas producing X-rays in the soft band [0.5 - 2.0 keV] \citep{Chadayammuri2022a,comparat_erosita_2022,Zhang_2024a,Zhang_2024b}. 

 
Most simulations are in poor agreement with current measurements from eROSITA \citep{Chadayammuri2022a,Zhang_2024a,Zhang_2024b,Silich2025} and there is currently no consensus among simulations replicating the X-ray emission from these observations \citep{Grayson2025}. Additionally, the recent tension between these X-ray stacked observations and thermal Sunyaev\textemdash Zel'dovich (tSZ) measurements of pressure measurements in similar galaxies makes it clear that there is still work to do before we fully understand the connection between feedback processes and the hottest gas phases in the CGM \citep[See][and the forthcoming \textit{X-rays in {\sc Romulus25} III}, Sanchez et al. 2026c, \textit{in prep}]{Bregman2022,Das2024,Ponnada2026,Zou2026}. 


With ongoing missions\textemdash including SRG/eROSITA and the upcoming ATHENA\textemdash the X-ray emitting gas within the CGM is a critical frontier through which we can improve our understanding of feedback, hot gas dynamics, and their impact on the overall evolution of galaxies. 


The first paper in the \textit{X-rays in {\sc Romulus25}} series, Sanchez et al. (2026a) \textit{submitted} (hereafter {\sc Paper I}) compares X-ray luminosity measurements from the cosmological {\sc Romulus25} simulation to observations from eROSITA. {\sc Paper I} shows that across most stellar mass bins, the simulated galaxy halos are brighter in soft band X-rays than observations. Only at L$_*$ and sub-L$_*$ masses do the mock X-ray observations fall well within observational errors motivating further exploration in this mass range. 

In this second paper of the \textit{X-rays in {\sc Romulus25}} series, we provide an in-depth and focused exploration of the X-ray emitting gas in the simulated Milky Way-mass regime. We explore the impact of varying gas properties such as density, temperature, and metallicity to further our understanding of X-ray emitting hot gas in the CGM. Additionally, we compare the luminosity profiles of individual genetically modified galaxies from the {\sc Pioneer} simulation suite\footnote{\citep[previously studied in][and run with the same physics as {\sc Romulus25}]{Sanchez2019,Butsky2020,Sanchez2021,Cruz2021,Hoelscher2026}} in an effort to identify potentially unique physics impacting the presence of hot, X-ray emitting gas in these halos. 

This paper is organized as follows: in Section 2, we briefly re-introduce the pertinent simulation details for the selected galaxy sample from {\sc Romulus25} and introduce the fully-cosmological PIONEER zoom-in suite of genetically modified galaxies; in Section 3, we describe our methods for calculating an total integrated X-ray surface brightness for each individual galaxy CGM; in Section 4, we describe the results from the PIONEER simulation analysis and use them to interpret the X-ray luminosity trends seen in {\sc Romulus25} and {\sc Paper I}. Finally, in Section 5, we discuss the broader implications for these results and summarize in Section 6.


\begin{table}[]
    \caption{PIONEER GM Galaxies}
    \centering
    \begin{tabular}{c c c c}
        \hline\hline 
        Sim  &  Halo Mass  & Stellar Mass & CGM Gas Mass \\
        Name & $M_{halo}/M_{\odot}$ & $M_{*}/M_{\odot}$  & ($M_{CGM}/M_{\odot}$) \\ 
        [0.5ex]
        \hline 
        R0  & 9.9 $\times$ 10$^{11}$ & 5.0 $\times$ 10$^{10}$ &  9.3 $\times$ 10$^{10}$  \\ 
        GM1 & 9.7 $\times$ 10$^{11}$ & 4.7 $\times$ 10$^{10}$ & 8.5 $\times$ 10$^{10}$   \\ 
        GM2 & 8.1 $\times$ 10$^{11}$ & 1.4 $\times$ 10$^{10}$ & 6.9 $\times$ 10$^{10}$   \\ 
        GM3 & 6.6 $\times$ 10$^{11}$ & 1.1 $\times$ 10$^{10}$ & 5.1 $\times$ 10$^{10}$   \\ 
        \hline
        R0 noBH  & 9.8 $\times$ 10$^{11}$ & 7.9 $\times$ 10$^{10}$ & 7.5 $\times$ 10$^{10}$   \\ 
        GM1 noBH & 9.9 $\times$ 10$^{11}$ & 7.4 $\times$ 10$^{10}$ & 8.0 $\times$ 10$^{10}$   \\ 
        GM2 noBH & 9.6 $\times$ 10$^{11}$ & 7.0 $\times$ 10$^{10}$ & 8.0 $\times$ 10$^{10}$   \\ 
        GM3 noBH & 8.4 $\times$ 10$^{11}$ & 7.3 $\times$ 10$^{10}$ & 6.4 $\times$ 10$^{10}$  \\ [1ex] 
    \end{tabular}
    \label{table:GM_table}
    \vspace{5mm}
\end{table}

\begin{figure}[t]
\centerline{\resizebox{1.05\hsize}{!}{\includegraphics[angle=0]{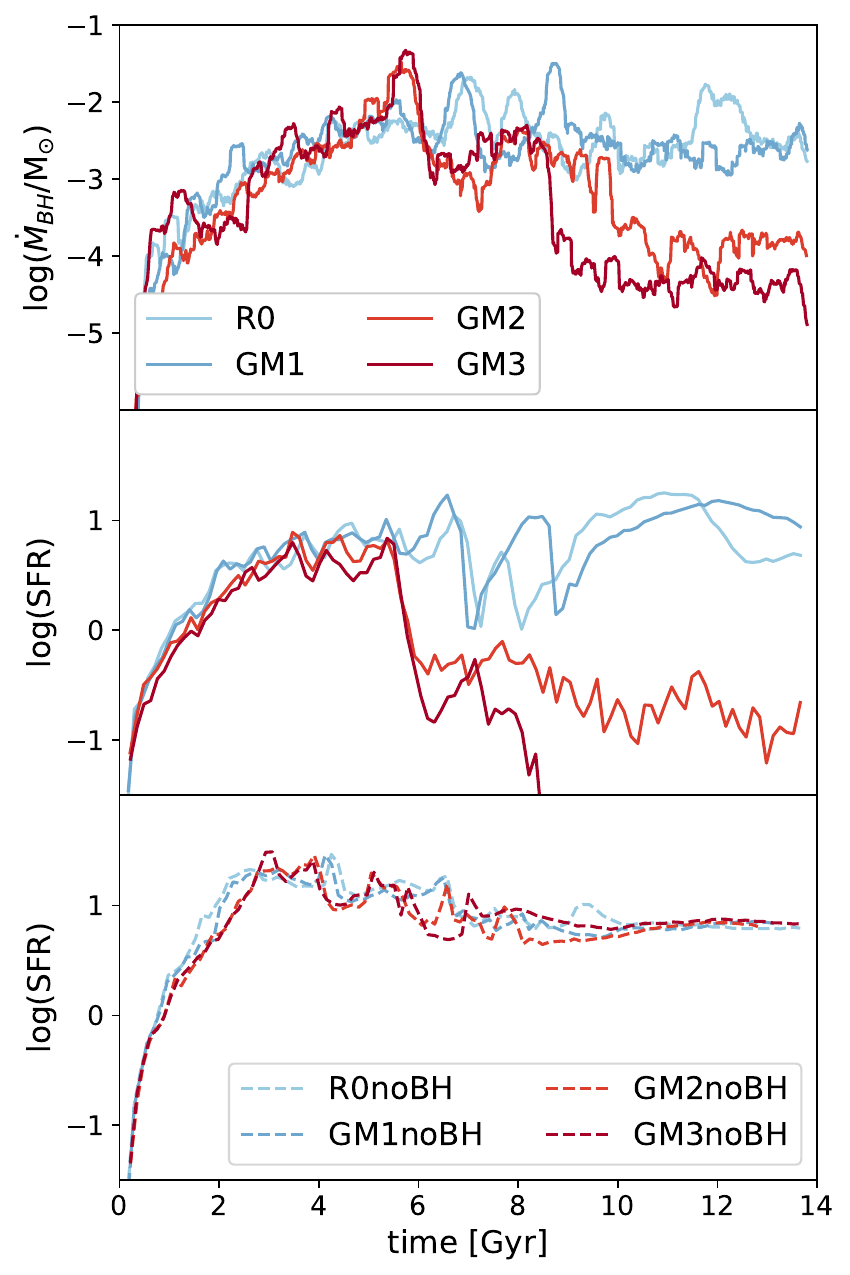}}}
\caption[]{The black hole accretion and star formation histories for the PIONEER simulations. (\textit{Upper}) The accretion history for the central SMBH for each of the four GM galaxies which includes two galaxies, R0 and GM1 (solid blue lines) which steadily accreted material into their SMBH until z = 0. The other two galaxies, GM2 and GM3 experience a peak in SMBH accretion at z $\sim$ 1, accrete steadily for 2 to 4 Gyr before having accretion fall steeply remain significantly low until z = 0. (\textit{Middle}) The SFH of the same GM galaxies above with BHs. Here, we see that R0 and GM1 remain star forming through their evolution to z = 0, while GM2 and GM3 are fully quenched of star formation at z = 1 and remain passive through to z = 0. (\textit{Lower}) The SFH for the four GM galaxies without BHs (dashed lines) all remain star forming throughout their evolution to z = 0. \\ 
}
\label{figure:SFHs}
\end{figure}

\section{Simulated Galaxy Sample}
\label{simulations}

The {\sc Romulus25} cosmological volume and the suite of genetically modified (GM) galaxies were all run with the Charm N-body GrAvity solver \citep[ChaNGa,][]{Menon2015}. Each simulation was run with a $\Lambda CDM$ cosmology used the following cosmological parameters: $\Omega _m$=0.3086, $\Omega _\Lambda$=0.6914, h=0.67, and $\sigma _8$=0.77 \citep{Planck2015}. The Plummer equivalent force softening lengths for each of these simulations is 250 pc, and a UV background from \cite{Haardt2012} is applied at z $\sim$ 9.

\begin{figure*}
\centerline{\resizebox{1\hsize}{!}{\includegraphics[angle=0]{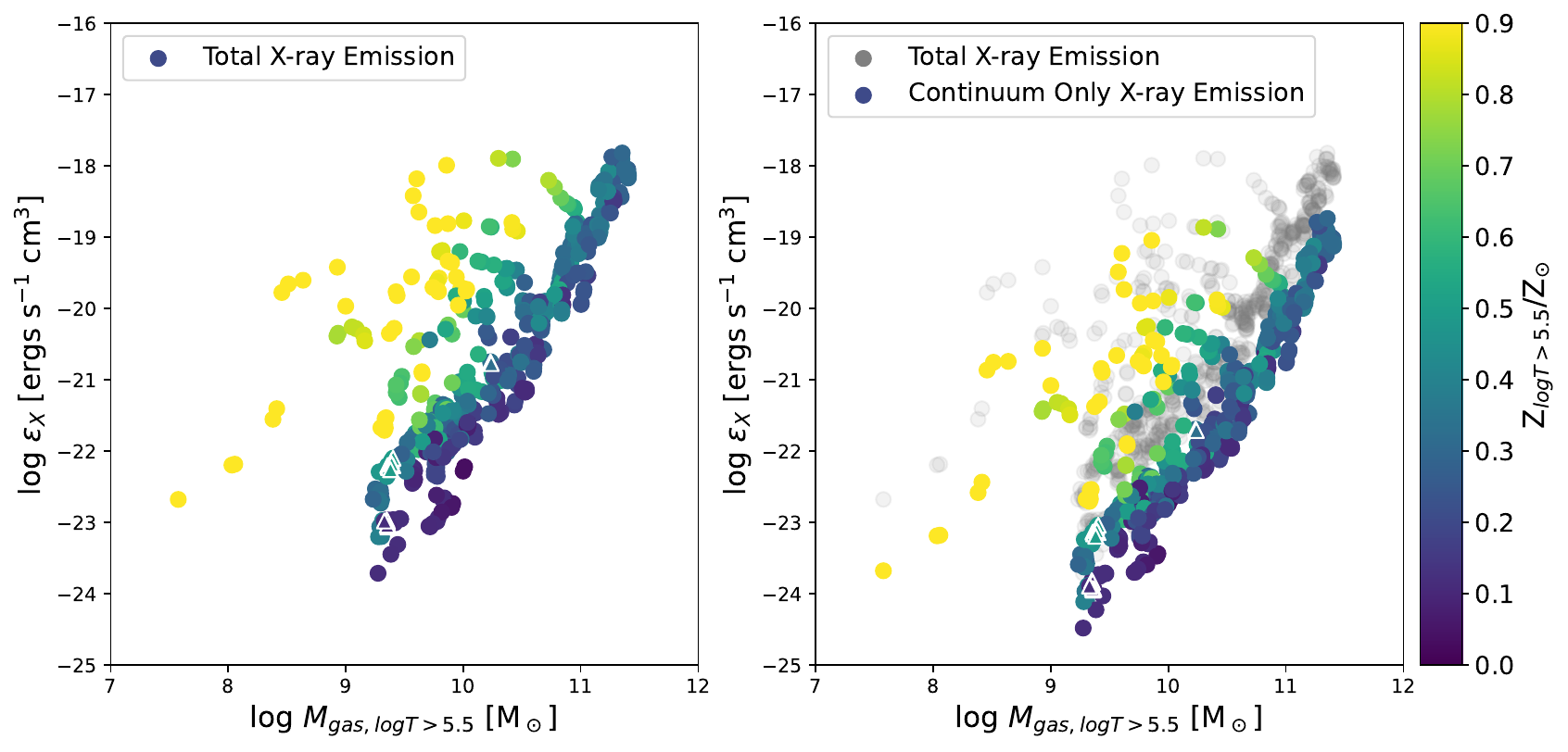}}}
\caption[]{Total X-ray emissivity of each galaxy halo in the Rekar subsample (calculated via an APEC model as described in Section \ref{xray-modeling}) with continuum and metal line emission (\textit{Left}) and only continuum emission (\textit{Right}) as a function of total hot gas (T \textgreater 10$^{5.5}$ K) in the CGM. Grey points on the right figure reproduce the metal line emission points from left figure.  Points are colored by average metallicity of the hot gas in the CGM. Circles indicate all galaxies which include a SMBH at its center, triangles with white outlines do not have a central SMBH. We find that the strong trend between x-ray emissivity in the soft band [0.5-2 keV] and the total amount of hot gas is preserved with or without the inclusion of metal line emission, as is the secondary trend with metallicity.}
\vspace{+2mm}
\label{figure:Ex_Mhotgas_Z}
\end{figure*}

Both suites used in this paper include all the same sub-grid physics including metal diffusion \citep{Shen2010}, star formation via Kroupa IMF \citep{Kroupa1993}, `blastwave' stellar feedback \citep{Stinson2006}, and BH accretion, feedback, and dynamical frictions prescriptions \citep{Tremmel2017}. As described in {\sc Paper I}, these simulations do not include a full metal line cooling treatment, instead using a low temperature extension to the metal cooling prescription which allows gas below T\textless$10^4$ K to cool proportionally to metals in the gas. See {\sc Paper I} for additional details including context, motivation, and a discussion of the impact of this limitation within the context of these results. We address the impact of no metal line cooling in the context of these results in Section \ref{sec-discussion}.

\subsection{X-ray Sample: Rekar from Romulus25}

The cosmological simulation {\sc Romulus25} is a 25 Mpc on-a-side volume which includes 1000 galaxies between log$_{10} (M_{*}/M_{\odot}$) = 6 - 11.5 at z = 0. The mass resolution of gas and dark matter particles are 2.1 $\times 10^5$ M$_{\odot}$ and 3.4 $\times 10^5$ M$_{\odot}$, respectively. We direct the reader to {\sc Paper I} for additional details on this simulation. 

In {\sc Paper I}, we introduce a sample of simulated galaxies from {\sc Romulus25} selected to match the stellar mass range from \cite{Zhang_2024a,Zhang_2024b} and call it the {\sc Rekar} sample. The observed galaxy samples of \cite{Zhang_2024b} are selected from the DESI legacy survey ninth data re-
lease (LS DR9) and completed ROSITA allsky sur-
vey (eRASS:4). The observations include over 200,000
galaxies and, in {\sc Paper I}, we compare to measurements of total integrated X-ray luminosity exclusively within their ''isolated galaxies`` sample. As in {\sc Paper I}, the galaxies in the simulated {\sc Rekar} sample are all between log$_{10} (M_{*}/M_{\odot}$) = 10 - 11.5 (Additional details including a mass histogram of this sample are available in Sanchez et al. 2026a, \textit{submitted}).

Throughout this paper, we will refer to measurements from {\sc Rekar} for both motivation and cosmological context within which to consider the results from the zoom-in runs of the {\sc Pioneer} suite.

\subsection{The PIONEER Simulations}

First introduced in \cite{Sanchez2019}, the cosmological {\sc Pioneer} simulation suite includes 4 zoom-in Milky-Way mass galaxy simulations run to z = 0. The resolution of these simulations are similar, though slightly higher than R25, and include a dark matter particle mass of 1.4 $\times 10^5$ M$_{\odot}$ and gas particle mass of 2.1 $\times 10^5$ M$_{\odot}$.

The {\sc Pioneer} simulations include the same BH physics models as R25. These simulations are characterized by nearly identical initial conditions with the exception of a minor merger modification which suppresses the mass of an infalling satellite at z = 1 (See Table \ref{table:GM_table}).  This genetic modification results in a set of galaxies that are characterized by varying star formation and galaxy-scale gas outflows driven by the SMBH \citep[For more details see][]{Sanchez2021}. Given the controlled nature of the genetic modification technique, we use these simulations as a controlled experiment for the effect of star formation rate and SMBH impact on the X-ray component of the CGM. Furthermore, this analysis also includes the same four {\sc Pioneer} simulations run without BH physics.

Figure \ref{figure:SFHs} shows the star formation histories (SFH) of the PIONEER suite split into the sample with and without BHs in the upper and lower panels, respectively. Blue solid lines indicate the star forming galaxies, R0 and GM1, while red/orange solid lines indicate the passive and quenched galaxies, GM2 and GM3. In the center panel, we denote the SFH from galaxies with no BHs with dashed line while the colors using the same color scheme: blue for ROnoBH and GM1noBH, and red/orange for GM2noBH and GM3noBH. Additionally, we note that all the galaxies with no BHs remain star forming throughout their evolution in the absence of black hole feedback \citep{Pontzen2017,Sanchez2019}. \\


\section{Forward Modeling Integrated X-ray Luminosity }
\label{xray-modeling}

To calculate the X-ray luminosity and surface brightness (SB) of the CGM around the galaxies, we define an exterior radius of $R_{500}$, the radius at which $\rho$ = 500 $\times$ $\rho_{crit}$, and an inner radius for the disk at 0.15$R_{500}$. For the rest of the analysis, we define $M_{CGM}$ as all the gas within these two boundaries: 0.15$R_{500}$ \textless $R_{CGM}$ \textless $R_{500}$.

As in {\sc Paper I}, we define the quantity, $L_{\chi}$, as the total luminosity of each simulated galaxy CGM, either from {\sc R25} or the {\sc Pioneer} Suite. We use the the predicted density, temperature, and metallicity from each galaxy's CGM gas and input these quantities into the APEC model from XSPEC \citep{XSPEC} to calculate the X-ray emission of each particle in the [0.5-2.0] keV soft X-ray band, $\epsilon_{\chi}$. We then calculate the total integrated X-ray luminosity of each galaxy as follows:
\begin{equation}
    L_{\chi} = \Sigma \epsilon_{\chi} n_H n_e V_{gas} 
\end{equation}
where n$_{H}$ is the hydrogen number density, n$_{e}$ is the electron number density, and V$_{gas}$ is the volume for each gas particle which is not set explicitely and is therefore calculated as V$_{gas}$ = m$_{gas}$/$\rho_{gas}$. 

From the X-ray luminosity of each particle, we calculate a soft X-ray surface brightness profile for the gas surrounding each galaxy using the publicly available smoothed-particle hydrodynamics (SPH) analysis code, {\sc pynbody} \citep{pynbody}. We note that unlike in the mock observations described in {\sc Paper I}, this S$_{\chi}$ calculation  \textit{does not} include any observational effects such as the eROSITA instrument response or galactic absorption. This method is consistent with measurements of this value from other simulations compared to in \citep{Zhang_2024b}. We calculate the X-ray SB profile from the luminosity using the equation:
\begin{equation}
    S_{\chi,bin} =  \frac{L_{\chi,bin}}{A_{bin}}
\end{equation}
where L$_{\chi,bin}$ is the sky-projected X-ray luminosity summed in each bin and A$_{bin}$ is the area of the spherical shell of the bin defined as $\pi(r_{outer}^2 - r_{inner}^2)$.


\begin{figure}[t!]
\vspace{5mm}
\centerline{\resizebox{1\hsize}{!}{\includegraphics[angle=0]{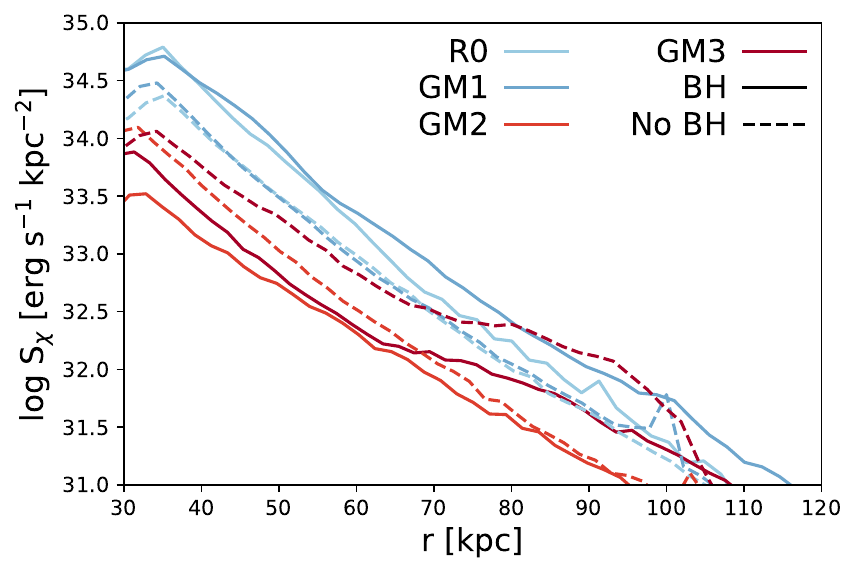}}}
\caption[]{The X-ray surface brightness profile of the 8 PIONEER GM simulated galaxies. Galaxies with and without BHs are denoted by solid and dashed lines, respectively. The two galaxies that are brightest in the X-ray soft band are R0 and GM1 (solid blue lines), which a have CGM that are impacted by both AGN and stellar feedback. Next in brightness are R0noBH and GM1noBH (blue dashed lines) and then GM2noBH and GM3noBH (red/orange dashed lines) which all have CGM that only have energy injected into them by stellar feedback. Notably, R0noBH and GM1noBH have larger stellar masses resulting in larger energy injection and brighter $S_{\chi}$ profiles. Finally, the two galaxies with the dimmest X-ray surface brightness profiles are GM2 and GM3 which each have central SMBH but have been fully quenched of both stellar and SMBH feedback since z = 1.}
\label{figure:GM_SBprof}
\vspace{2mm}
\end{figure}

\section{Results}
\label{sec-results}

\subsection{Physical Drivers of X-ray Brightness}

In {\sc Paper I}, we focused on measuring mock SB profiles and total X-ray luminosities for all of the {\sc Rekar} galaxy halos from R25 to compare with observations \citep{Chadayammuri2022a,Zhang_2024a,Zhang_2024b}. Here, we begin by exploring the specific gas phase properties in the hot CGM that give rise to X-ray brightness. 

In Figure \ref{figure:Ex_Mhotgas_Z}, we find a strong trend, both with (\textit{left panel}) and without (\textit{right panel}) metal line emission,  between the X-ray emission in the soft X-ray band and the total amount of hot gas (log(T/K) \textgreater 5.5) in the CGM, M$_{gas,logT >5.5}$. Even without including metal line emission, this strong trend is prevalent across nearly three orders of magnitude in gas mass, log M$_{gas,logT >5.5}$ = 9 \textemdash 12, and across the X-ray emissivity from -24 to -19. 

In both panels, we also find the presence of a strong secondary trend conveyed through the color bar where the average metallicity of the hot gas, $\bar{Z}_{gas,logT >5.5}$, appears to increase the X-ray emissivity in the soft band [0.5 - 2.0] keV for the same amount of M$_{gas,logT >5.5}$. Motivated by these correlations and given the similar prescriptions used by both R25 and the galaxies in the GM controlled experiment, the next sections focus on the GM galaxies to identify the underlying physics resulting in the presence of hot gas mass in the CGM and process through which  metal enrichment changes the X-ray brightness. 

\subsection{Total Feedback sets Hot Gas Mass}
\label{hotgasmass}

 We begin by showing the profiles of X-ray surface brightness and emissivity in the CGM of each GM galaxy in Figure \ref{figure:GM_SBprof}. Here and for the rest of the analysis, we focus on the continuum-only X-ray emission and subsequent luminosity calculations as they are consistent with these simulations  \cite[][Sanchez et al. 2026a, \textit{submitted}]{Sanchez2019} which lack a full metal line cooling treatment and are more consistent with observations from \cite{Zhang_2024b}. 

In Figure \ref{figure:GM_SBprof}, we show the soft X-ray surface brightness of each GM galaxy as a function of distance from the galaxy center. All eight GM galaxies in the sample show a bright central profile with a shallow peak at about 30 kpc (due to the removal of the galaxy disk) and a decreasing profile from about 40 kpc beyond 100 kpc. A distinct trend emerges when we compare the galaxy halos with highest surface brightness from brightest to dimmest. The two galaxies that are brightest in the soft X-ray band (within the inner $\sim$ 70 kpc) are those that remain actively star forming and have an active AGN present through to z = 0 (Figure \ref{figure:SFHs}). Next in surface brightness, the galaxies without black holes follow roughly in order of stellar mass: R0noBH, GM1noBH, GM3noBH, GM2noBH. We note that GM3noBH shows a flatter profile, maintaining a brighter halo at larger radii (\textgreater 50 kpc) compared to the others. Finally, the galaxies with the lowest inner-halo surface brightness are the galaxies with black holes which have been quenched by z = 1: GM2 and GM3. Similar to GM3noBH, the surface brightness profile of GM3 also is also flatter beyond $\sim$60 kpc and has a brighter outer halo compared to GM2.

Thus, a trend becomes apparent: galaxies with ongoing stellar and BH activity appear brightest in soft band X-rays, followed by galaxies with ongoing stellar activity, and dimmest are the galaxies with no ongoing activity (and therefore no ongoing energy injection into the gas). From this figure, neither stellar feedback nor AGN feedback from the SMBH appear more significant in driving soft band X-ray surface brightness, implying that the total energy injection into the halo sets the X-ray brightness. We explore this in more detail below. 

\begin{figure}
\centerline{\resizebox{1\hsize}{!}{\includegraphics[angle=0]{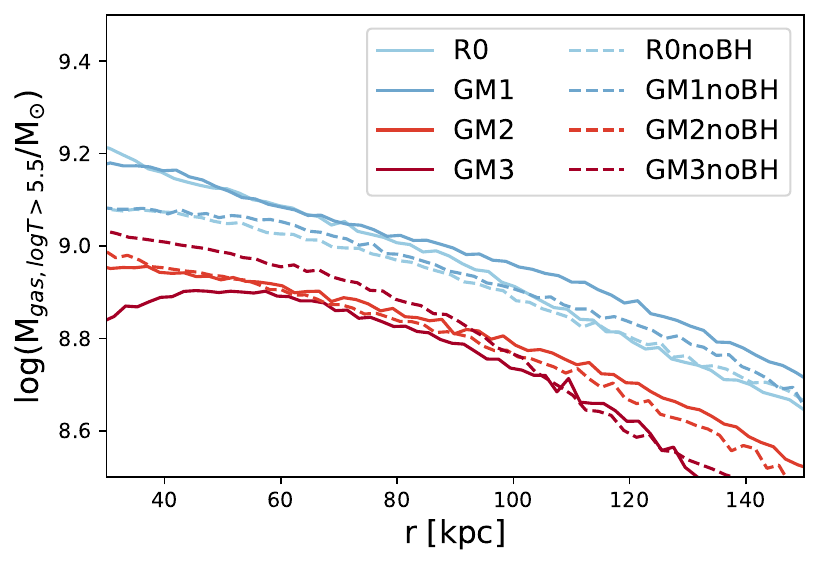}}}
\caption[]{Profiles of the hot gas mass (T \textgreater 10$^{5.5}$ K) in the CGM for all eight of the GM galaxies. Colors and linestyles as in Figure \ref{figure:GM_SBprof}. We find a similar trend in this figure as we see when examining the X-ray surface brightness profiles (Figure \ref{figure:GM_SBprof}). The galaxies that are the brightest in X-rays are the galaxies with the most hot gas mass in their CGM: P0 and GM1. Next in order of hot gas mass are the pairs, P0noBH and GM1noBH, and then, GM2noBH and Gm3noBH. Finally, the galaxies with the least amount of hot gas mass (and similarly the least X-ray emission) are GM2 and GM3, the only two quenched galaxies at z=0.}
\label{figure:GM_hotgasmass}
\vspace{+2mm}
\end{figure}

Not surprisingly, we see a similar trend when we examine the amount of hot gas in the CGM of the galaxies, M$_{gas,logT >5.5}$. Figure \ref{figure:GM_hotgasmass} shows the total hot gas mass (log(T/K) \textgreater 5.5) of each galaxy CGM as a function of distance from the galaxy center in $\sim$ 2 kpc bins. As in Figure \ref{figure:GM_SBprof}, the order of most to least amount of hot gas mass follows as: the galaxies with both ongoing stellar and BH feedback, R0 and GM1; galaxies with no black holes but ongoing stellar feedback, R0noBH and GM1noBH which both have the larger stellar masses of the noBH galaxies; galaxies with no black holes but ongoing stellar feedback and lower stellar masses, GM2noBH and GM3noBH; and finally GM2 and GM3, both of which have no ongoing feedback activity at z = 0.

The presence of this trend in terms of hot gas mass appears to be similarly driven by the feedback energy injection from both stellar and AGN activity which results in the presence of this hot, X-ray emitting gas in the halo. 

\begin{figure}
    \centering
    \includegraphics[width=1\linewidth]{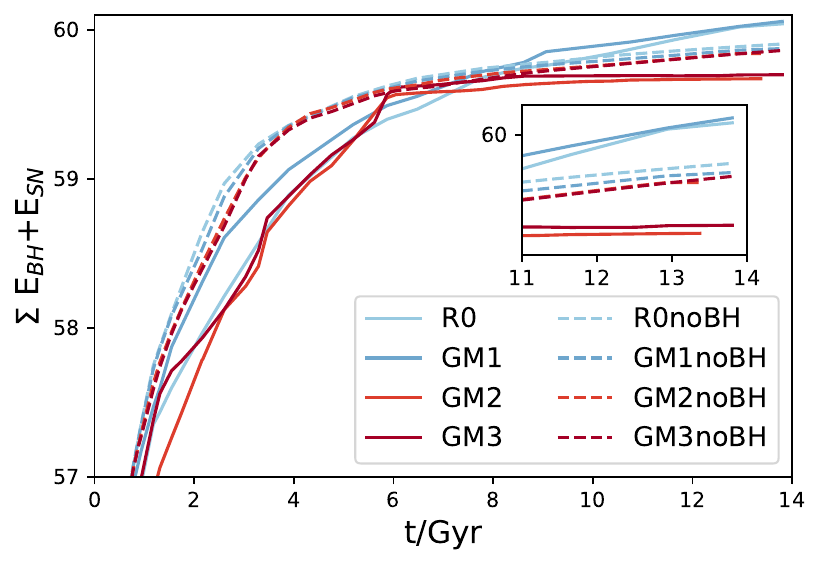}
    \caption{Total energy injection from both SMBH feedback and stellar feedback in all of the 8 GM galaxies. Colors and linestyles as in Figure \ref{figure:GM_SBprof}. At z = 0, the galaxies with the most total energy injection from both stars and the SMBH are R0 and GM1, the same galaxies that are also brightest in X-ray surface brightness. Next in terms of total energy injection are the galaxies with no black hole physics, approximately in order of stellar mass which correlates to stellar feedback. Lastly, as in X-ray surface brightness, GM2 and GM3 have the least energy injection having been quenched at z $\sim$ 1. By comparing this figure with Figure \ref{figure:GM_hotgasmass}, we can directly tie the total energy injected by AGN and stellar feedback to the presence of hot gas mass in the CGM and in turn the X-ray SB.}
    \label{fig:totenergyinj}
    \vspace{2mm}
\end{figure}

In Figure \ref{fig:totenergyinj}, we confirm the trend we noted in the previous two plots: the total energy injection by both stellar and AGN feedback determines the amount of hot gas mass, M$_{gas,logT>5.5}$, present in the CGM. Neither individual process appears to dominate the soft X-ray band luminosity, instead the cumulative sum of total energy injected into the CGM over the evolution of the galaxy appears to set this property. This result indicates that X-ray observations may be unable to isolate the impact of stellar and/or AGN feedback in the CGM; we discuss the implications of this result further below. 

\subsection{Metal-Rich Hot Gas is Both Hotter and Brighter}

Alongside the strong emissivity and hot gas mass trend show in Figure \ref{figure:Ex_Mhotgas_Z}, we note again the prominent secondary trend with the average metallicity of the hot gas in the CGM (log(T/K) \textgreater 5.5). We find that galaxies with the same amount of hot gas in the CGM will have much brighter X-ray emission in the soft band \textit{if the hot gas is metal rich}. 

\begin{figure}
\centerline{
\resizebox{1\hsize}{!}{\includegraphics[angle=0]{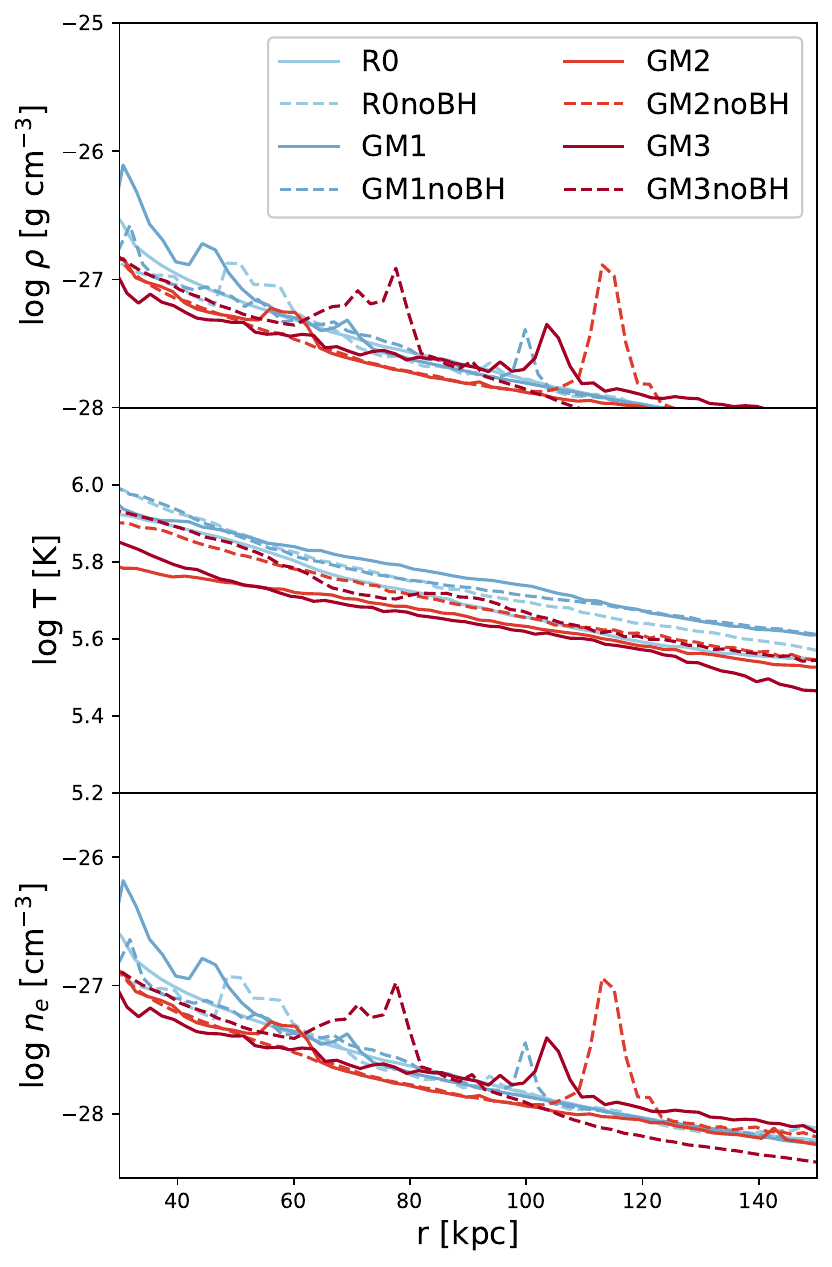}}}
\caption[]{The profiles of gas density (\textit{Upper}), temperature (\textit{Middle}), and electron density (\textit{Lower}) for the CGM of the eight GM galaxies. Line styles and colors as in Figure \ref{figure:GM_SBprof}. Unlike in the profiles of X-ray SB and hot gas mass, the profiles of temperature, gas density, and electron density do not show the same trends as the previous figures and the measurements in each set of profiles are much more consistent. There are slight deviations: for example, the inner CGM temperature of the two quenched galaxies GM2 and GM3 is a bit lower than the rest which are all star forming. There are also a few discrete peaks in the density profiles at varying distances in the CGM due to the presence of dwarf galaxy satellites. However, none of these properties are significantly distinct enough to explain the differences in X-ray brightness in the GM galaxy sample.}
\label{figure:GM_profiles}
\end{figure}

To better understand how the presence of metal-rich hot gas increases X-ray emissivity more than pristine hot gas, we explore the physical characteristics of the hot gas present in the CGM of the PIONEER simulations at z = 0. 

\begin{figure}[]
\centerline{\resizebox{1\hsize}{!}{\includegraphics[angle=0]{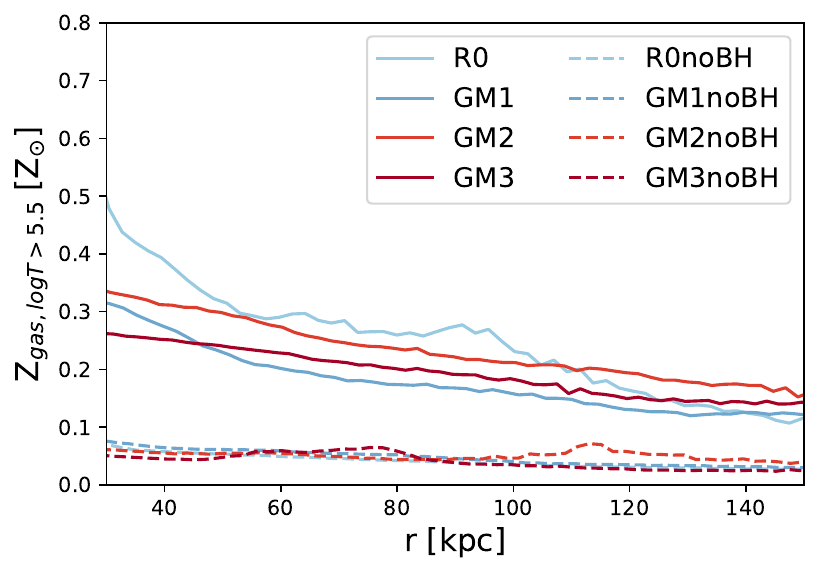}}}
\caption[]{Metallicity profiles for the eight GM galaxy CGM. Colors and linestyles as in Figure \ref{figure:GM_SBprof}. There is a distinct bimodality shown in this figure between the galaxies with BHs (solid lines) which all have similarly high metallicities and those without BHs (dashed lines) which all have roughly an order of magnitude lower metallcity. One exception is GM2noBH which has a slightly higher metallicity profile beyond about 120 kpc. This difference is likely due to metal enrichment from dwarf galaxies in the outer CGM, which also explains the small bump in the metallicity profile of GM3noBH at $\sim$ 70 kpc. In addition to the presence of hot gas mass in the CGM, the difference in the metallicities of these galaxies with and without black holes also appears to play a significant role in setting the X-ray brightness of their CGM.}
\vspace{3mm}
\label{figure:GM_metals}
\end{figure}

Figure \ref{figure:GM_profiles} shows the radial profiles of gas density (\textit{Top panel}), temperature (\textit{Middle panel}), and electron density (\textit{Bottom panel}) for the hot CGM gas of the 8 PIONEER GM galaxies. Colors and linestyles as in Figure \ref{figure:GM_SBprof}. While the differences in the surface brightness profiles reach up to an order of magnitude between the brightest (GM1) and dimmest (GM2), the three properties in this figure do not significant differences either between the different GMs or their noBH counterparts. However, we see a very significant difference between the total metallicities of the PIONEER galaxies. Figure \ref{figure:GM_metals} shows the strong bimodality between the galaxies with and without BHs. Galaxies with BHs have nearly an order of magnitude more metals in the CGM than the galaxies without BHs. \cite{Sanchez2019} determined that this stark difference was due to the SMBH in the GMs (solid lines) ejecting metals from the central galaxy out into the CGM. However, AGN driven metal enrichment in the CGM appears to be limited to the MW-mass regime; more recently, measurements of OVI in the CGM of galaxies across a wide range of stellar mass (log M$_{*}$ = 7 - 11.5) indicate that below M$_{*}$ \textless 10$^{9.5}$ M$_{\odot}$ BH feedback is not the dominant process evacuating metals from the disk (Rubalcava et al. 2026, \textit{in prep}).  

We expand on this result further in the next section by exploring these characteristics in the hot gas halos of galaxies from the {\sc Rekar} sample from {\sc Romulus}.\\

\subsection{Cosmological Implications for {\sc Romulus25}}

By exploring the {\sc Pioneer} simulations in this way as a controlled experiment for the impact of feedback on CGM gas, it appears that the critical differences between these galaxies in terms of the X-ray emitting gas are: (1) the total amount of hot gas mass ($T_{gas}$ \textgreater $10^{5.5}$ K) present and (2) the average metallicity of this quantity of hot gas.

While the relationship between energy injection, hot gas mass, and X-ray emissivity is self consistent\textemdash more energy injection results in more hot gas which subsequently results in more X-ray emission\textemdash the explanation for the secondary metallicity relationship is more interesting. Despite its higher metallicity, the X-ray emitting hot gas does not show a significant difference in electron density, a potential first-order explanation (Figure \ref{figure:GM_profiles}). Instead, we find that the increased metallicity increases the overall charge of the gas ions. This increased charge then accelerates the associated electrons at higher rates which in turn changes the overall scale and slope of the energy spectrum across the soft X-ray band [0.5-2.0] keV from Brehmsstralung radiation.

To show this explicitly, we calculate the X-ray emissivity of the hot gas in the CGM (only through free-free/Brehmsstralung radiation) using the following equation:

\begin{equation}
    \epsilon_{ff} = 6.8 \times 10^{-38} Z^2 n_e n_i T^{1/2} e^{hv/kT} \bar{g}_{ff}
\end{equation}
where Z is the charge, $n_e$ is the electron density, $n_i$ is the number density of ions, and $g_{ff}$ is the Gaunt factor. Here, we use a frequency and velocity averaged Gaunt factor between the range 1.1 to 1.5; as the Gaunt factor is a constant and we compare these same measurements between the galaxies, the specific choice does not impact our results. While the Gaunt factor may also be temperature-dependent, this difference varies at most by factor of $\sim$2, with $g_{ff}$ varying between 1 to 2 \citep{deAvillez2015} and we argue that these differences are negligible between the galaxies given their substantial differences in emissivity.

\begin{figure*}
\centerline{\resizebox{.99\hsize}{!}{\includegraphics[angle=0]{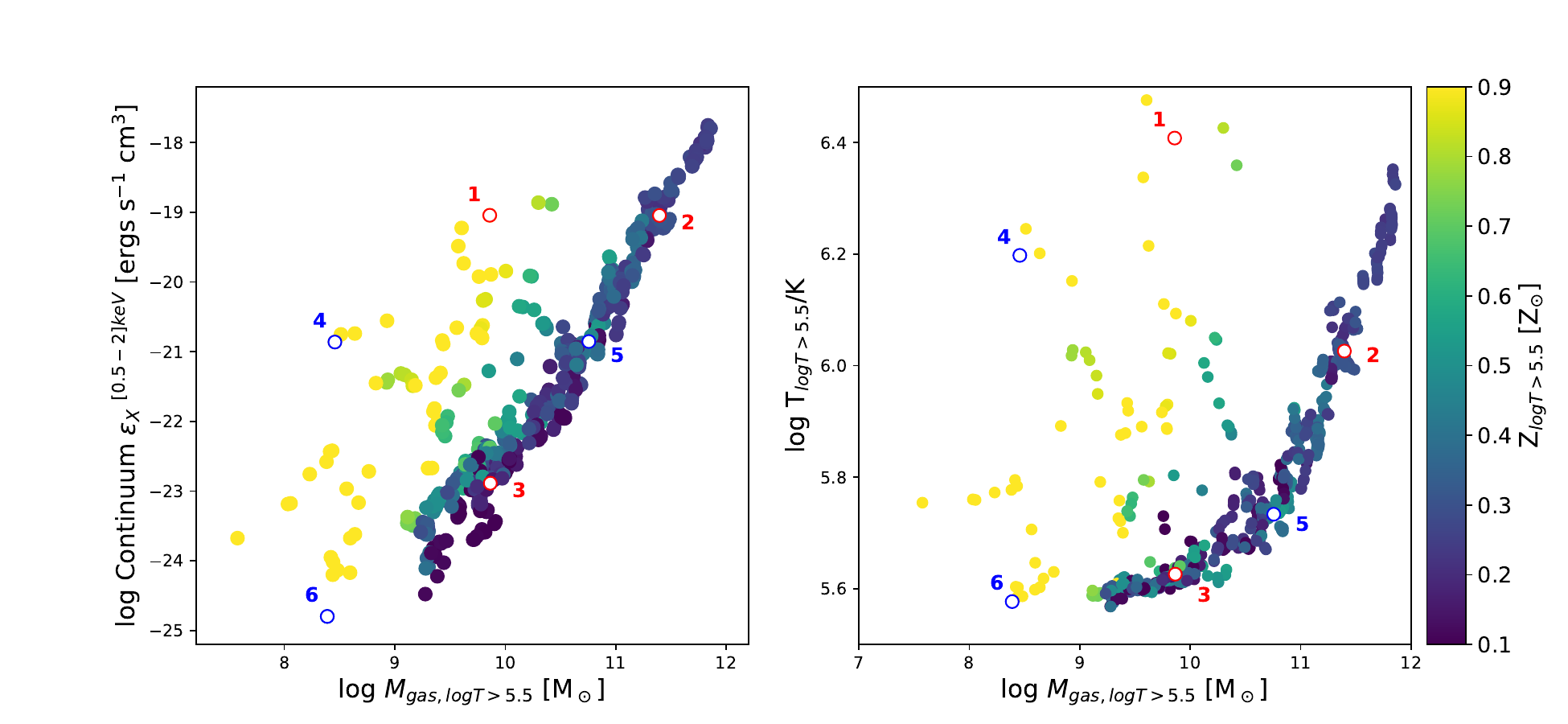}}}
\caption[]{(\textit{Left:}) Soft band X-ray emissivity for the galaxies from the {\sc Romulus25} Rekar sample introduced in {\sc Paper I} as a function of the hot gas mass in their CGM. Points are colored by the average hot gas mass metallicity and individual galaxies have been hand-selected (white dots) and labeled (numbers 1 through 7) for further exploration in the right panel. There is a strong linear trend between the soft X-ray emission and total hot gas mass in the CGM and a secondary trend with the metallicity of the hot gas. (\textit{Right:}) Average overall temperature of the hot gas, log(T/K) > 5.5), as a function of hot gas mass as in the left panel. Unlike in the GM case, in the {\sc Romulus25} sample, galaxies with metal-rich CGM are also hotter as a result of the feedback energy both heating gas and moving metals of out the disk.}
\label{figure:intLx_Mhotgas_brehm}
\vspace{2mm}
\end{figure*}

We show the impact of different metallicities on the photon energy in Figure \ref{figure:intLx_Mhotgas_brehm} using the {\sc Rekar} sample of galaxies from {\sc Romulus25} (see {\sc Paper I} Simulation Parameters for more details). In the left panel, we again show the X-ray emissivity as a function of hot CGM gas mass colored by average metallicity of the hot CGM gas (as in Figure \ref{figure:Ex_Mhotgas_Z}, right panel). Here, we highlight two hand-selected galaxy trios (1,2,3 labeled in red \& 4,5,6 labeled in blue) that display a combination of comparable X-ray emissivity, average metallicities, and/or hot gas mass. 

In the right panel, we show the average temperature of the hot gas mass in the CGM similarly as a function of the hot CGM gas mass. As expected, the average temperature of the hot gas has a strong trend with total hot gas mass, and in bins of similar hot gas mass, metal-enriched gas is hotter. It follows that gas that has experienced the most heating through feedback would also the most metal-rich. 

We also see that our example galaxies show significant differences between average temperature in this hot phase gas, in order from highest to lowest average T they are: 1, 4, 2, 5, 3, and 6. Unlike in the GM case where this difference is just about 0.2 dex, as we do see a distinguishable difference in temperature between these example galaxies, we must take both metallicity and temperature into consideration when examining the impact on total continuum emission.

\begin{figure}
    \centering
    \includegraphics[width=0.95\linewidth]{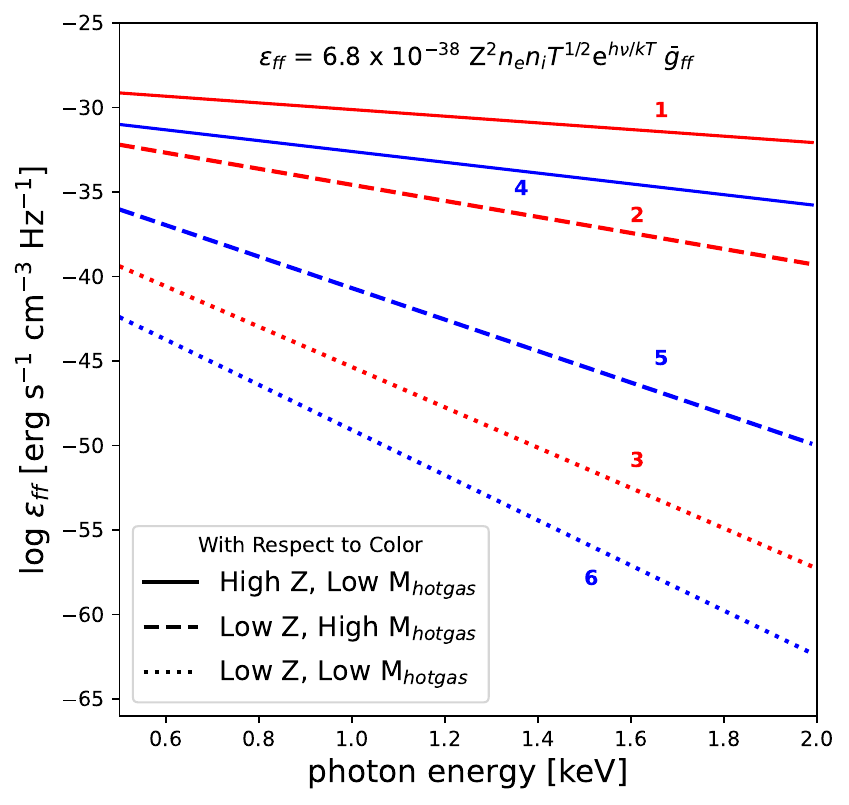}
    \caption{Soft X-ray spectrum for the hand-selected galaxies shown in Figure \ref{figure:intLx_Mhotgas_brehm}. Galaxies with higher average metallicity CGM are also hotter and tend to have higher and flatter emission spectra, while galaxies with less hot gas mass have steeper emission spectra and are not emitting as much X-ray energy.}
    \label{fig:keV}
\end{figure}

In Figure \ref{fig:keV}, we plot the soft X-ray band spectrum of the six selected galaxies as a function of photon energy; each line is colored to match their corresponding selected trio in red or blue, and denoted by differing linestyles as described below.

\begin{table}[]
    \caption{R25 Selected Comparison Sample in Figure \ref{figure:intLx_Mhotgas_brehm}}
    \centering
    \begin{tabular}{c c c c}
        \hline\hline 
        Galaxy  &  Halo Mass  & Stellar Mass & Hot CGM Gas Mass \\
        Label & $M_{halo}/M_{\odot}$ & $M_{*}/M_{\odot}$  & ($M_{gas,logT>5.5}/M_{\odot}$) \\ 
        [0.5ex]
        \hline 
        1  & 5.7 $\times$ 10$^{11}$ & 7.8 $\times$ 10$^{10}$ & 7.2  $\times$ 10$^{9}$  \\ 
        2 & 3.4 $\times$ 10$^{12}$ & 7.1 $\times$ 10$^{10}$ &  2.5 $\times$ 10$^{11}$   \\ 
        3 & 5.3 $\times$ 10$^{11}$ & 2.0 $\times$ 10$^{10}$ & 7.3 $\times$ 10$^{9}$   \\ 

        \hline
        4  & 3.0 $\times$ 10$^{11}$ & 3.9 $\times$ 10$^{10}$ &  2.9 $\times$ 10$^{8}$   \\ 
        5 & 1.1 $\times$ 10$^{12}$ & 2.5 $\times$ 10$^{10}$ & 5.7 $\times$ 10$^{10}$   \\ 
        6 & 1.8 $\times$ 10$^{11}$ & 1.0 $\times$ 10$^{10}$ &  2.5 $\times$ 10$^{8}$   \\ [1ex] 
    \end{tabular}
    \label{table:R25_table}
    \vspace{2mm}
\end{table}

By comparing between these paired sets of galaxies, we see the distinct differences in the spectrum due to different hot gas masses and its varying average temperature and metallicity in the CGM. Exploring the first galaxy trio (1,2,3, in red), we find :
\begin{itemize}
    \item \textit{Galaxies 1 \& 2}: Galaxy 1 has a higher average temperature and metallicity, $\bar{Z}_{logT>5.5}$, but a smaller amount of hot gas in its CGM compared to Galaxy 2. Nevertheless, the soft X-ray spectrum of Galaxy 1 peaks above $\epsilon _{ff}$ = -30 erg s$^{-1}$ cm$^{-3}$ Hz$^{-1}$ and has a flatter, negative slope (red solid line). Meanwhile, the spectrum of Galaxy 2 peaks at only $\epsilon _{ff}$ $\eqsim$ -33 erg s$^{-1}$ cm$^{-3}$ Hz$^{-1}$ (red dashed line). Despite over an order of magnitude difference in hot gas mass, Galaxies 1 and 2 each emit roughly the same total amount of soft band X-ray emission.
    \item \textit{Galaxies 1 \& 3}: In this pair, Galaxy 1 still has the higher average temperature and metallicity but the same amount of hot gas mass compared to Galaxy 3. The spectrum of Galaxy 3 peaks at $\epsilon _{ff}$ $\sim$ 38 erg s$^{-1}$ cm$^{-3}$ Hz$^{-1}$ and has a steeply declining slope (red dotted line), while Galaxy 1's spectrum is much flatter and peaks much higher at $\epsilon _{ff}$ $\sim$ -29 erg s$^{-1}$ cm$^{-3}$ Hz$^{-1}$ (red solid line). Galaxies 1 and 3, despite the same quantity of hot gas mass, emit significantly different amounts of emission (nearly 5 orders of magnitude difference) in the X-ray soft band.

\end{itemize}

In this first trio, a galaxy with less hot gas in the CGM but higher average temperature and metallicity in its hot gas component can emit the same amount of total X-ray emission as a galaxy with unenriched but significantly more hot gas in the CGM (up to 2 orders of magnitude). This increase in overall soft X-ray emissivity is a result of the increased intensity and flattened slope of the soft X-ray spectrum that results from the hotter, metal-rich gas in galaxy 1.

We repeat the analysis with a second trio, similarly hand-selected to match via hot gas mass, average hot gas metallicity, and/or X-ray emissivity.
\begin{itemize}
    \item \textit{Galaxies 4 \& 5}: Galaxy 4 contains nearly two orders of magnitude less hot gas mass in its CGM compared to Galaxy 5, but the former has a much higher average temperature and metallicity which results in both galaxies being equally bright in the soft X-ray band. Similar to Galaxy 1, Galaxy 4 has a mostly flat spectrum, though gently negative and decreasing in slope (and also lower in intensity compared to Galaxy 1). Meanwhile, Galaxy 5 has spectrum with  a sharper decline from its peak at $\epsilon _{ff}$ $\simeq$ -35 erg s$^{-1}$ cm$^{-3}$ Hz$^{-1}$.
    \item \textit{Galaxies 4 \& 6}: Finally, Galaxies 4 \& 6 contain nearly identical amounts of hot gas mass in the CGM, but the hot gas in Galaxy 4 is much more metal rich. As a result, the spectrum of Galaxy 4 peaks at $\epsilon _{ff}$ $\simeq$ -29 erg s$^{-1}$ cm$^{-3}$ Hz$^{-1}$, while the spectrum of Galaxy 6 peaks much lower at $\epsilon _{ff}$ $\simeq$ -42 erg s$^{-1}$ cm$^{-3}$ Hz$^{-1}$. The slope of Galaxy 6 is also much steeper as it decreases. 
\end{itemize}

In this second trio, the galaxy that is brighter in X-ray emission is similarly the galaxy with the more metal-rich hot gas in the CGM as in the first trio. 

In these two trios, the galaxy pairs 1 \& 4, 2 \& 5, and 3 \& 6 are mirrors of each other in their selection. However, we note that while pairs 1 \& 4 and 3 \& 6 have the same slope/shape to their soft X-ray spectrum, the Galaxies 2 \& 5 are more different which is also true for their temperatures. Not only is the intensity/peak of Galaxy 2's X-ray emissivity higher (again, by selection) but the slope of the spectrum is different, and flatter than in Galaxy 5. While the metallicity does increase the overall X-ray emissivity in the soft band, we find the differences in the slopes of the emission spectra show that the average temperature of hot gas mass remains the more dominant component in setting the X-ray emissivity. Furthermore, gas at T = 10$^{6}$ K (kT $\sim$ 0.01 keV) sits at the tail of the X-ray soft band spectrum which is why such small differences in temperature can make a large impact on the overall emission spectra. 

By analyzing these two sets of galaxy trios with varying metal enrichment, average temperature, and total CGM quantity of hot gas present, as well as the subsequent X-ray emissivity, we confirm that the results from the GM controlled experiment hold in a cosmological context and better understand the relationship between these physical drivers in setting X-ray brightness in simulated galaxies from {\sc Romulus25}.



\section{Discussion}
\label{sec-discussion}

\subsection{Constraining Feedback through Panchromatic CGM Studies}
In our study, we find that soft X-ray emission does not depend strongly on only one type of feedback, either from the central SMBH or from stars uniquely. Instead, the total energy injection into the gas of the CGM sets the total X-ray brightness in the soft band. Though we determine that feedback energy impacts the galaxy by heating and evacuating gas into the CGM (whether metal-rich or not), we can not use soft X-ray emission directly to differentiate between varying individual feedback sources.

However, it may be possible to use a combination of CGM observations to potentially differentiate between feedback sources impacting individual galaxy CGM. \cite{Sanchez2019} looked at this same suite of {\sc PIONEER} GM galaxies and determined that the SMBH was key to driving metal-rich gas out of the disk and into the CGM at Milky Way-masses. They also determined that measurements from OVI directly traced the metal enrichment which resulted from the central SMBH; in other words, L* galaxies with more significant AGN feedback are shown to produce higher OVI column densities. 

Therefore, it may be possible to observe a sample of unique L* galaxy candidates using both UV absorption spectroscopy and soft X-ray emission to determine which galaxy CGM are characterized by multiple sources of feedback or SMBH feedback alone in the MW-mass regime. This result could significantly constrain one facet of the complicated evolution history of individual 
L* galaxies. Furthermore, by identifying individual galaxies that have experienced specific types of feedback conditions, we may be able to set constraints on other key characteristics of these galaxies, such as distinct outflow rates, dominant CGM phases, and more.

While current X-ray missions are not yet capable of measuring the X-ray emission of individual local galaxy CGM, upcoming missions like Athena or a high-spectral-resolution microcalorimeter (like the recently passed Line Emission Mapper, LEM) may make this kind of comparison a soon-to-be reality. 

\subsection{Metal-Rich Gas May Bias X-ray Observations}

This study is the first to explore the impact of metallicity on the hot, X-ray emitting gas component of the CGM to connect it to the unique physics driving X-ray luminosity of a gas halo. Prior analyses of X-ray bright clusters and groups assume more pristine gas reservoirs where the impact of metallicity or how it may change X-ray emission has not yet been constrained.

This paper motivates future observations to break the degeneracy between CGM reservoirs with high amounts of pristine hot gas mass from lower mass reservoirs that are hotter and metal-rich. Our study also indicates that bright metal-line emission studies of X-ray ions (such as O VII, O VIII, Ne IX, and Fe XVI) may be biased towards high metallicity CGM due to the required presence of a metal-rich halos to observe these high ions \citep{Nicastro2018,Mathur2023,Wang2026}. \cite{Schellenberger2024}, while exploring metal-line emission studies in support of the LEM probe concept, note a similar degeneracy between X-ray bright gas density and metallicity in the simulated galaxies of Illustris, EAGLE, and SIMBA.

Finally, in broad strokes, new and ongoing observations in this hot gas regime, like SRG/eROSITA and the upcoming ATHENA, will be key as they provide more robust constraints for cosmological simulations. In this still under-studied phase of gas (i.e. the very hot CGM), further comparisons to observations are key in helping us determine if current sub-grid models are effectively modeling this hottest gas phase of a galaxy halo. 



\section{Summary and Conclusion}
\label{sec-conclude}

This paper is the second in a series exploring the hot, X-ray bright phase of gas in galaxy CGM from the {\sc Romulus25} simulation. In {Paper I}, we compare measurements from these simulated galaxies to observations from eROSITA and determine that while the simulations are characterized by an overall overabundance of hot gas in their CGM, we well-match observations within the Milky Way-mass regime.

Here in {\sc Paper II}, we explore the physical properties of hot soft X-ray [0.5-2.0] keV emitting gas in the halo of galaxies, with a focus on the Milky Way-mass regime. By exploring the hot CGM component of galaxies from {\sc Romulus25} and the genetically modified {\sc PIONEER} simulations, we determine two main characteristics of the X-ray emitting gas phase in a galaxy halo that determine overall X-ray brightness. 

Our main takeaways include:
\begin{enumerate}
    \item X-ray emission and luminosity in the soft X-ray band [0.5-2.0] keV of our simulated galaxy CGM are set by two main gas properties: 1) total amount of hot gas mass (log (T/K) \textgreater 5.5) and 2) the average temperature and metallicity of that mass of hot gas. 
    \item The total amount of hot gas mass (M$_{gas,logT>5.5}$) is set by the total combined energy injection of both stellar and AGN feedback contributed to the CGM over the evolution of the galaxy. Neither stellar feedback or feedback from the SMBH is more dominant in setting the quantity of M$_{gas,logT>5.5}$.
    \item As a result of total feedback energy setting the amount of hot gas mass, galaxies with the hottest gas are also metal-rich. Feedback processes can both heat gas and move metals out of the disk resulting in these hotter, metal-rich halos. 
    \item The average metallicity of M$_{gas,log(T/{\rm K})>5.5}$ is a second, and highly important characteristic of the X-ray bright gas. A galaxy with a metal-rich hot gas phase will be significantly brighter in soft X-rays than a galaxy with the same quantity of hot gas mass that is more pristine. The metal rich hot gas in these halos is brighter because the increased charge of the ions in the gas changes the intensity of the spectrum across the X-ray soft band range. 
    \item While our analysis is focused on the MW-mass regime, we confirm that the impact of these two physical properties in setting the total X-ray emissivity of a galaxy halo holds across the full stellar mass range of our {\sc Romulus25} galaxy sample: $log_{10}(M_{*}/M_{\odot})$ = 10 - 11.5.
\end{enumerate}

These results indicate that the average temperature and metallicity as well as quantity of hot (log T \textgreater 5.5) gas in the CGM directly sets the X-ray brightness in the soft band [0.5-2.0] keV. These results hold for galaxies from the cosmological volume across a stellar mass range of $log_{10}(M_{*}/M_{\odot})$ = 10 - 11.5. Paper III (Sanchez et al. 2026c, \textit{in prep}) will further expand on these findings. The next paper will measure the thermal pressure (via thermal Sunyaev Zeldovich) of the hot gas halos of the suite of {\sc Romulus25} galaxies, the GM {\sc PIONEER} suite, as well as a similar simulation with cosmic rays to compare them with our previous X-ray luminosity measurements and address the recent observational disparity between CMB measurements and X-ray observations in the local universe.


\section{Acknowledgements}

This work was supported by the NSF MPS-Ascend award ID 2212959. 
NNS acknowledges useful conversations with Lia Medeiros, Skylar Grayson, Johan Comparat, Annalisa Pillepich, Yi Zhang, Soumya Shreeram, Sam Wu, Sam Ponneda, and Cameron Hummels.

BDO's contribution was supported by the Grant ATR2025-164580 funded by MICIU/AEI/10.13039/50110001103.

Part of this work was performed in part at Aspen Center for Physics, which is supported by NSF grant PHY-2210452.

The simulation in this study was run on the NAS HPC computer Pleaides, Blue Waters, and TACC Stampede2. Analysis was completed on NAS HEC SMD-21-75544133 and NASA HEC SMD-16-7592.

This research was supported by the Munich Institute for Astro-, Particle and BioPhysics (MIAPbP) which is funded by the Deutsche Forschungsgemeinschaft (DFG, German Research Foundation) under Germany´s Excellence Strategy – EXC-2094 – 390783311.

Any opinions, findings, and conclusions or recommendations expressed in this material are those of the author(s) and do not necessarily reflect the views of the National Science Foundation.

\bibliography{Sanchez2025b_XrayPhys.bib}

\begin{thebibliography}{}
\expandafter\ifx\csname natexlab\endcsname\relax\def\natexlab#1{#1}\fi

\bibitem[{{Alexander} {et~al.}(2025){Alexander}, {Hickox}, {Aird}, {Combes}, {Costa}, {Habouzit}, {Harrison}, {Leng}, {Morabito}, {Uckelman}, \& {Vickers}}]{Alexander2025}
{Alexander}, D.~M., {Hickox}, R.~C., {Aird}, J., {et~al.} 2025, 101, 101733

\bibitem[{{Arnaud}(1996)}]{XSPEC}
{Arnaud}, K.~A. 1996, in Astronomical Society of the Pacific Conference Series, Vol. 101, Astronomical Data Analysis Software and Systems V, ed. G.~H. {Jacoby} \& J.~{Barnes}, 17

\bibitem[{{Bregman} {et~al.}(2022){Bregman}, {Hodges-Kluck}, {Qu}, {Pratt}, {Li}, \& {Yun}}]{Bregman2022}
{Bregman}, J.~N., {Hodges-Kluck}, E., {Qu}, Z., {et~al.} 2022, \apj, 928, 14

\bibitem[{{Butsky} {et~al.}(2020){Butsky}, {Fielding}, {Hayward}, {Hummels}, {Quinn}, \& {Werk}}]{Butsky2020}
{Butsky}, I.~S., {Fielding}, D.~B., {Hayward}, C.~C., {et~al.} 2020, \apj, 903, 77

\bibitem[{{Chadayammuri} {et~al.}(2022){Chadayammuri}, {Bogd{\'a}n}, {Oppenheimer}, {Kraft}, {Forman}, \& {Jones}}]{Chadayammuri2022a}
{Chadayammuri}, U., {Bogd{\'a}n}, {\'A}., {Oppenheimer}, B.~D., {et~al.} 2022, \apjl, 936, L15

\bibitem[{{Ciotti} {et~al.}(2010){Ciotti}, {Ostriker}, \& {Proga}}]{Ciotti2010}
{Ciotti}, L., {Ostriker}, J.~P., \& {Proga}, D. 2010, \apj, 717, 708

\bibitem[{Comparat {et~al.}(2022)Comparat, Truong, Merloni, Pillepich, Ponti, Driver, Bellstedt, Liske, Aird, Brüggen, Bulbul, Davies, Villalba, Georgakakis, Haberl, Liu, Maitra, Nandra, Popesso, Predehl, Robotham, Salvato, Thorne, \& Zhang}]{comparat_erosita_2022}
Comparat, J., Truong, N., Merloni, A., {et~al.} 2022, Astronomy \& Astrophysics, 666, A156

\bibitem[{{Crenshaw} \& {Kraemer}(2012)}]{Crenshaw2012}
{Crenshaw}, D.~M., \& {Kraemer}, S.~B. 2012, \apj, 753, 75

\bibitem[{{Crenshaw} {et~al.}(2003){Crenshaw}, {Kraemer}, {Gabel}, {Kaastra}, {Steenbrugge}, {Brinkman}, {Dunn}, {George}, {Liedahl}, {Paerels}, {Turner}, \& {Yaqoob}}]{Crenshaw2003}
{Crenshaw}, D.~M., {Kraemer}, S.~B., {Gabel}, J.~R., {et~al.} 2003, \apj, 594, 116

\bibitem[{{Cruz} {et~al.}(2021){Cruz}, {Pontzen}, {Volonteri}, {Quinn}, {Tremmel}, {Brooks}, {Sanchez}, {Munshi}, \& {Di Cintio}}]{Cruz2021}
{Cruz}, A., {Pontzen}, A., {Volonteri}, M., {et~al.} 2021, \mnras, 500, 2177

\bibitem[{{Das} {et~al.}(2024){Das}, {Chiang}, {Allen}, {Mathur}, \& {Gupta}}]{Das2024}
{Das}, S., {Chiang}, Y.-K., {Allen}, S., {Mathur}, S., \& {Gupta}, A. 2024, in EAS2024, 59

\bibitem[{Davies {et~al.}(2022)Davies, Pontzen, \& Crain}]{davies_galaxy_2022}
Davies, J.~J., Pontzen, A., \& Crain, R.~A. 2022, 14, 1, arXiv: 2203.08157

\bibitem[{{de Avillez, Miguel A.} \& {Breitschwerdt, Dieter}(2015)}]{deAvillez2015}
{de Avillez, Miguel A.}, \& {Breitschwerdt, Dieter}. 2015, A\&A, 580, A124

\bibitem[{{Fabian}(2012)}]{Fabian2012}
{Fabian}, A.~C. 2012, \araa, 50, 455

\bibitem[{{Grayson} {et~al.}(2025){Grayson}, {Scannapieco}, {Comparat}, {ZuHone}, {Zhang}, {Shreeram}, {Br{\"u}ggen}, \& {Bulbul}}]{Grayson2025}
{Grayson}, S., {Scannapieco}, E., {Comparat}, J., {et~al.} 2025, \apj, 994, 89

\bibitem[{Haardt \& Madau(2012)}]{Haardt2012}
Haardt, F., \& Madau, P. 2012, The Astrophysical Journal, 746, 125

\bibitem[{{Hoelscher} {et~al.}(2026){Hoelscher}, {Holley-Bockelmann}, {Cruz}, \& {Sanchez}}]{Hoelscher2026}
{Hoelscher}, Z.~J., {Holley-Bockelmann}, K., {Cruz}, A., \& {Sanchez}, N.~N. 2026, \apj, 1004, 222

\bibitem[{{Kormendy} \& {Ho}(2013)}]{Kormendy2013}
{Kormendy}, J., \& {Ho}, L.~C. 2013, \araa, 51, 511

\bibitem[{{Kroupa} {et~al.}(1993){Kroupa}, {Tout}, \& {Gilmore}}]{Kroupa1993}
{Kroupa}, P., {Tout}, C.~A., \& {Gilmore}, G. 1993, \mnras, 262, 545

\bibitem[{{Mathur} {et~al.}(2023){Mathur}, {Das}, {Gupta}, \& {Krongold}}]{Mathur2023}
{Mathur}, S., {Das}, S., {Gupta}, A., \& {Krongold}, Y. 2023, \mnras, 525, L11

\bibitem[{Menon {et~al.}(2015)Menon, Wesolowski, Zheng, Jetley, Kale, Quinn, \& Governato}]{Menon2015}
Menon, H., Wesolowski, L., Zheng, G., {et~al.} 2015, Computational Astrophysics and Cosmology, 2, 1

\bibitem[{{Mestici} {et~al.}(2024){Mestici}, {Tombesi}, {Gaspari}, {Piconcelli}, \& {Panessa}}]{Mestici2024}
{Mestici}, S., {Tombesi}, F., {Gaspari}, M., {Piconcelli}, E., \& {Panessa}, F. 2024, \mnras, 532, 3036

\bibitem[{{Nelson} {et~al.}(2018){Nelson}, {Kauffmann}, {Pillepich}, {Genel}, {Springel}, {Pakmor}, {Hernquist}, {Weinberger}, {Torrey}, {Vogelsberger}, \& {Marinacci}}]{Nelson2018}
{Nelson}, D., {Kauffmann}, G., {Pillepich}, A., {et~al.} 2018, \mnras, 477, 450

\bibitem[{{Nicastro} {et~al.}(2018){Nicastro}, {Kaastra}, {Krongold}, {Borgani}, {Branchini}, {Cen}, {Dadina}, {Danforth}, {Elvis}, {Fiore}, {Gupta}, {Mathur}, {Mayya}, {Paerels}, {Piro}, {Rosa-Gonzalez}, {Schaye}, {Shull}, {Torres-Zafra}, {Wijers}, \& {Zappacosta}}]{Nicastro2018}
{Nicastro}, F., {Kaastra}, J., {Krongold}, Y., {et~al.} 2018, \nat, 558, 406

\bibitem[{{Oppenheimer} {et~al.}(2020){Oppenheimer}, {Bogd{\'a}n}, {Crain}, {ZuHone}, {Forman}, {Schaye}, {Wijers}, {Davies}, {Jones}, {Kraft}, \& {Ghirardini}}]{Oppenheimer2020}
{Oppenheimer}, B.~D., {Bogd{\'a}n}, {\'A}., {Crain}, R.~A., {et~al.} 2020, \apjl, 893, L24

\bibitem[{{Padovani} {et~al.}(2017){Padovani}, {Alexander}, {Assef}, {De Marco}, {Giommi}, {Hickox}, {Richards}, {Smol{\v{c}}i{\'c}}, {Hatziminaoglou}, {Mainieri}, \& {Salvato}}]{Padovani2017}
{Padovani}, P., {Alexander}, D.~M., {Assef}, R.~J., {et~al.} 2017, \aapr, 25, 2

\bibitem[{{Panessa} {et~al.}(2019){Panessa}, {Baldi}, {Laor}, {Padovani}, {Behar}, \& {McHardy}}]{Panessa2019}
{Panessa}, F., {Baldi}, R.~D., {Laor}, A., {et~al.} 2019, Nature Astronomy, 3, 387

\bibitem[{{Piacitelli} {et~al.}(2025){Piacitelli}, {Brooks}, {Christensen}, {Sanchez}, {Faerman}, {Shen}, {Cruz}, {Keller}, {Quinn}, \& {Wadsley}}]{Piacitelli2025}
{Piacitelli}, D.~R., {Brooks}, A.~M., {Christensen}, C., {et~al.} 2025, \apj, 993, 230

\bibitem[{{Piacitelli} {et~al.}(2026){Piacitelli}, {Brooks}, {Sanchez}, {Khatri}, {Christensen}, {Hummels}, {Mishra}, {Cruz}, {Keller}, {Quinn}, {Shen}, \& {Wadsley}}]{Piacitelli2026}
{Piacitelli}, D.~R., {Brooks}, A.~M., {Sanchez}, N.~N., {et~al.} 2026, arXiv e-prints, arXiv:2605.15275

\bibitem[{{Planck Collaboration}(2016)}]{Planck2015}
{Planck Collaboration}. 2016, Astronomy and Astrophysics, 594, arXiv:1502.01589

\bibitem[{{Ponnada} {et~al.}(2026){Ponnada}, {Hopkins}, {Lu}, {Silich}, {Butsky}, \& {Kere{\v{s}}}}]{Ponnada2026}
{Ponnada}, S.~B., {Hopkins}, P.~F., {Lu}, Y.~S., {et~al.} 2026, \apjl, 997, L13

\bibitem[{{Pontzen} {et~al.}(2013){Pontzen}, {Ro{\v s}kar}, {Stinson}, {Woods}, {Reed}, {Coles}, \& {Quinn}}]{pynbody}
{Pontzen}, A., {Ro{\v s}kar}, R., {Stinson}, G.~S., {et~al.} 2013, {pynbody: Astrophysics Simulation Analysis for Python}, astrophysics Source Code Library, ascl:1305.002

\bibitem[{{Pontzen} {et~al.}(2017){Pontzen}, {Tremmel}, {Roth}, {Peiris}, {Saintonge}, {Volonteri}, {Quinn}, \& {Governato}}]{Pontzen2017}
{Pontzen}, A., {Tremmel}, M., {Roth}, N., {et~al.} 2017, \mnras, 465, 547

\bibitem[{{Roth} {et~al.}(2016){Roth}, {Pontzen}, \& {Peiris}}]{Roth2016}
{Roth}, N., {Pontzen}, A., \& {Peiris}, H.~V. 2016, \mnras, 455, 974

\bibitem[{Sanchez {et~al.}(2021)Sanchez, Tremmel, Werk, Pontzen, Christensen, Quinn, Loebman, \& Cruz}]{Sanchez2021}
Sanchez, N.~N., Tremmel, M., Werk, J.~K., {et~al.} 2021, The Astrophysical Journal, 911, 116

\bibitem[{Sanchez {et~al.}(2019)Sanchez, Werk, Tremmel, Pontzen, Christensen, Quinn, \& Cruz}]{Sanchez2019}
Sanchez, N.~N., Werk, J.~K., Tremmel, M., {et~al.} 2019, The Astrophysical Journal, 882, 8

\bibitem[{{Sanchez} {et~al.}(2024){Sanchez}, {Werk}, {Christensen}, {Telford}, {Quinn}, {Tremmel}, {Mead}, {Sharma}, \& {Brooks}}]{Sanchez2024}
{Sanchez}, N.~N., {Werk}, J.~K., {Christensen}, C., {et~al.} 2024, \apj, 967, 100

\bibitem[{{Schellenberger} {et~al.}(2024){Schellenberger}, {Bogd{\'a}n}, {ZuHone}, {Oppenheimer}, {Truong}, {Khabibullin}, {Jennings}, {Pillepich}, {Burchett}, {Carr}, {Chakraborty}, {Crain}, {Forman}, {Jones}, {Kilbourne}, {Kraft}, {Markevitch}, {Nagai}, {Nelson}, {Ogorzalek}, {Randall}, {Sarkar}, {Schaye}, {Veilleux}, {Vogelsberger}, {Wang}, \& {Zhuravleva}}]{Schellenberger2024}
{Schellenberger}, G., {Bogd{\'a}n}, {\'A}., {ZuHone}, J.~A., {et~al.} 2024, \apj, 969, 85

\bibitem[{Shen {et~al.}(2010)Shen, Wadsley, \& Stinson}]{Shen2010}
Shen, S., Wadsley, J., \& Stinson, G. 2010, Monthly Notices of the Royal Astronomical Society, 407, 1581

\bibitem[{{Silich} {et~al.}(2025){Silich}, {ZuHone}, {Bellomi}, {Hummels}, {Oppenheimer}, {Hopkins}, {Lochhaas}, {Ponnada}, \& {Vikhlinin}}]{Silich2025}
{Silich}, E.~M., {ZuHone}, J., {Bellomi}, E., {et~al.} 2025, \apj, 993, 125

\bibitem[{Stinson {et~al.}(2006)Stinson, Seth, Katz, Wadsley, Governato, \& Quinn}]{Stinson2006}
Stinson, G., Seth, A., Katz, N., {et~al.} 2006, Monthly Notices of the Royal Astronomical Society, 373, 1074

\bibitem[{{Suresh} {et~al.}(2026){Suresh}, {Blanton}, \& {Rennehan}}]{Suresh2017}
{Suresh}, A., {Blanton}, M.~R., \& {Rennehan}, D. 2026, \apj, 997, 105

\bibitem[{Tremmel {et~al.}(2017)Tremmel, Karcher, Governato, Volonteri, Quinn, Pontzen, Anderson, \& Bellovary}]{Tremmel2017}
Tremmel, M., Karcher, M., Governato, F., {et~al.} 2017, Monthly Notices of the Royal Astronomical Society, 470, 1121

\bibitem[{{Truong} {et~al.}(2021){Truong}, {Pillepich}, {Nelson}, {Werner}, \& {Hernquist}}]{Truong2021}
{Truong}, N., {Pillepich}, A., {Nelson}, D., {Werner}, N., \& {Hernquist}, L. 2021, \mnras, 508, 1563

\bibitem[{{Truong} {et~al.}(2020){Truong}, {Pillepich}, {Werner}, {Nelson}, {Lakhchaura}, {Weinberger}, {Springel}, {Vogelsberger}, \& {Hernquist}}]{Truong2020}
{Truong}, N., {Pillepich}, A., {Werner}, N., {et~al.} 2020, \mnras, 494, 549

\bibitem[{{Truong} {et~al.}(2023){Truong}, {Pillepich}, {Nelson}, {Bogd{\'a}n}, {Schellenberger}, {Chakraborty}, {Forman}, {Kraft}, {Markevitch}, {Ogorzalek}, {Oppenheimer}, {Sarkar}, {Veilleux}, {Vogelsberger}, {Wang}, {Werner}, {Zhuravleva}, \& {Zuhone}}]{Truong2023}
{Truong}, N., {Pillepich}, A., {Nelson}, D., {et~al.} 2023, \mnras, 525, 1976

\bibitem[{{Wang} {et~al.}(2026){Wang}, {Zhou}, {Fang}, \& {Nicastro}}]{Wang2026}
{Wang}, K., {Zhou}, Z., {Fang}, T., \& {Nicastro}, F. 2026, \apj, 1009, 40

\bibitem[{Zhang {et~al.}(2024{\natexlab{a}})Zhang, Comparat, Ponti, Meloni, Nandra, Haberl, Locatelli, Zhang, Sanders, Zheng, Liu, Popesso, Liu, Truong, Pillepich, Predehl, \& Salvato}]{Zhang_2024a}
Zhang, Y., Comparat, J., Ponti, G., {et~al.} 2024{\natexlab{a}}, The {Hot} {Circum}-{Galactic} {Medium} in the {eROSITA} {All} {Sky} {Survey} {I}. {X}-ray {Surface} {Brightness} {Profiles}, arXiv:2401.17308 [astro-ph]

\bibitem[{Zhang {et~al.}(2024{\natexlab{b}})Zhang, Comparat, Ponti, Meloni, Nandra, Haberl, Truong, Pillepich, Locatelli, Zhang, Sanders, Zheng, Liu, Popesso, Liu, Predehl, \& Salvato}]{Zhang_2024b}
---. 2024{\natexlab{b}}, The {Hot} {Circum}-{Galactic} {Medium} in the {eROSITA} {All} {Sky} {Survey} {II}. {Scaling} {Relations} between {X}-ray {Luminosity} and {Galaxies}' {Mass}, arXiv:2401.17309 [astro-ph]

\bibitem[{{Zou} {et~al.}(2026){Zou}, {Yuan}, {Ji}, {He}, {Li}, {Zhang}, {Comparat}, {Qu}, \& {Fang}}]{Zou2026}
{Zou}, Y., {Yuan}, F., {Ji}, S., {et~al.} 2026, \apj, 1003, 199

\end{thebibliography}

\end{document}